\documentclass[12pt]{amsart}

\normalbaselines
\newtheorem{theorem}{Theorem}[section]

\theoremstyle{definition}

\theoremstyle{remark}
\newtheorem{remark}[theorem]{Remark}
\theoremstyle{corollary}
\newtheorem{corollary}{Corollary}[section]
\numberwithin{equation}{section}
\newcommand{\abs}[1]{\lvert#1\rvert}

\newcommand{\N}{\mathcal{N}}
\newcommand{\B}{\mathfrak{B}}
\newcommand{\an}{\mathfrak{a}}
\newcommand{\n}{\mathfrak{n}}
\newcommand{\HI}{\mathfrak{H}}

\begin{document}
\title{Temporally stable Coherent states for a free magnetic Schr\"odinger operator }

\author{K. Thirulogasanthar$^{\dagger}$}
\author{Nasser Saad$^{\ddagger}$}
\author{Attila B. von Keviczky$^{\ast}$}
\address{$\dagger,\;\ast$\;Department of Mathematics and Statistics, Concordia University,
7141 Sherbrooke Street West, Montreal, Quebec H4B 1R6, Canada }
\email{$\dagger\;$santhar@vax2.concordia.ca}
\address{$\ddagger$ Department of Mathematics and Statistics, University of Prince Edward Island, 550 University avenue, Charlottetown, PEI, C1A 4P3, Canada.}
\email{$\ddagger$\;nsaad@upei.ca}

\date{\today}


\begin{abstract}
Eigenfunctions and eigenvalues of the free magnetic Schr\"odinger operator, 
describing a spinless particle confined to an infinite layer of fixed width,  
are discussed in detail. The eigenfunctions are realized as an orthonormal basis 
of a suitable Hilbert space.
Four different classes of temporally stable coherent states associated to the 
operator are presented. The first two classes are derived as coherent states with one
 degree of freedom and the last two classes are derived with two degrees of freedom. 
The dynamical algebra of each class is found. Statistical quantities associated to each 
class of coherent states are calculated explicitely.
\end{abstract}

\maketitle

\pagestyle{myheadings}

\markboth{K.Thirulogasanthar, N.~Saad, A.B.~von Keviczky }{Temporally stable CS for a magnetic operator}



\section{Introduction}
By generalizing the definition of canonical coherent states, CS for short, 
 Gazeau and Klauder \cite{Gk} proposed a method to construct temporally stable
 CS for a quantum system with one degree of freedom. Since then, the method 
has been successfully applied to different quantum systems \cite{Tem,F}. 
As an extension of \cite{Gk}, a method  was presented to build CS for systems 
with several degrees of freedom \cite{Ng}.
Motivated from the recent interest on temporally stable coherent states, we present 
in this article four different classes of CS using the spectrum of the free magnetic
 Schr\"odinger operator
\begin{equation}\label{s11}
H_0=\frac{1}{2M}(\text{\bf P}-\frac{e}{c}\text{\bf A})^2,
\end{equation}
where $\text{\bf A}$ is the magnetic vector potential, $e$ is the charge of the particle, $c$ is the speed of light, and $\text{\bf P}=-i\hbar\nabla$ with $\hbar$ being the Planck's constant divided by $2\pi$. By constructing CS for the operator $H_0$ we also demonstrate the method proposed in \cite{Ng} and analyze the temporal stability and action identity conditions for the multidimensional case. These features were excluded from the discussion of \cite{Ng}.\\
The article is organized as follows. In Section 2, we introduce the detailed description of the free magnetic Schr\"odinger operator (\ref{s11}), exploring its spectrum and the eigenvectors. In Section 3 we realize the eigenfunctions of Section 2 as an orthonormal basis of a Hilbert space. For the sake of completeness in Section 4 we discuss the definition of Gazeau-Klauder CS. In Section 5, associated with the spectrum of (\ref{s11}), two classes of CS with one degree of freedom are constructed. In Section 6, two classes of CS with two degrees of freedom are constructed. In Section 7, detail classification of the dynamical algebra is provided. In section 8 we explicitely calculate the quantum statistical quantities associated to the CS.


\section{the free magnetic Schr\"odinger operator}
\noindent Consider an infinite layer of fixed width $d$, that is, $\Sigma=\mathbb R^2\times[0,d]$. Suppose the layer is placed into a perpendicular homogeneous magnetic field of intensity $\text{\bf B}=(0,0,B)$, where $B$ is a constant. The Hamiltonian of this system can be written using (\ref{s11}) as
\begin{equation}\label{s21}
H_0=\frac{1}{2M}\left(\text{\bf P}^2-\frac{e}{c}\text{\bf P}\cdot\text{\bf A}-\frac{e}{c}\text{\bf A}\cdot\text{\bf P}+\frac{e^2}{c^2}\text{\bf A}^2\right).
\end{equation}
When the circular gauge $\text{\bf A}=-\frac{1}{2}\text{\bf r}\times \text{\bf B}=\frac{1}{2}|B|(-y,x,0)$ is chosen, we have
for a state vector $\psi$
$$\text{\bf P}\cdot\text{\bf A}\psi=-i\hbar(\nabla\cdot \text{\bf A})\psi-i\hbar\text{\bf A}\cdot\nabla\psi=\text{\bf A}\cdot\text{\bf P}\psi.$$
Consequently, a spinless quantum particle confined to the layer is described by the free magnetic Schr\"odinger operator
\begin{equation}\label{s22}
H_0=\frac{1}{2M}\left(\text{\bf P}^2-\frac{2e}{c}\text{\bf A}\cdot\text{\bf P}+\frac{e^2}{c^2}\text{\bf A}^2\right)
\end{equation}
acting in $L^2(\Sigma)$ with Dirichlet boundary conditions
$$\psi(\text{\bf x},0)=\psi(\text{\bf x},d)=0,\;\;\text{\bf x}=(x,y)\in\mathbb R^2.$$
In the absence of an additional interaction, the operator $H_0$ can be written as
\begin{equation}\label{s23}
H_0=-\frac{\hbar^2}{2M}\nabla^2+\frac{ie\hbar |B|}{2Mc}\left(x\frac{\partial}{\partial y}-y\frac{\partial}{\partial x}\right)+\frac{e^2|B|^2}{8Mc^2}(x^2+y^2).
\end{equation}
The presence of the potential $x^2+y^2$ suggests the use of the cylindrical coordinates for the separation of the variables. Thus we have
\begin{equation}\label{s24}
H_0=-\frac{\hbar^2}{2M}\nabla^2+\frac{e^2|B|^2}{8Mc^2}r^2+\frac{ie\hbar |B|}{2Mc}\frac{\partial}{\partial \theta}, \end{equation}
where
$$ \nabla^2=\frac{\partial^2}{\partial r^2}+\frac{1}{r}\frac{\partial}{\partial r}+\frac{1}{r^2}\frac{\partial^2}{\partial \theta^2}+\frac{\partial^2}{\partial z^2}.$$
If we define the cyclotron frequency $\omega_c=-\frac{e|B|}{Mc}$, then
\begin{equation}\label{s25}
H_0=-\frac{\hbar^2}{2M}\nabla^2+\frac{M\omega_c^2}{8}r^2+\frac{\omega_c }{2}L_z, \ \text{where } L_z=-i\hbar \frac{\partial}{\partial \theta}.
\end{equation}
Let
$$\Psi(r,\theta,z)=\psi(r,\theta)\chi(z),$$
we can easily find that the differential equation satisfied by $\chi(z)$ and obeying boundary conditions $\chi(0)=\chi(d)=0$ yields
\begin{equation}\label{s26}
\chi_n(z)\equiv \sqrt{\frac{2}{d}}\sin(\frac{\pi n z}{d}),\quad n=1,2,\dots
\end{equation}
which form an orthonormal basis in $L^2[0,d]$. Note that the case of $n=0$ correspondence to $\chi_0(z)=0$ is physically insignificant. The corresponding eigenvalues are 
\begin{equation}\label{s27}
\epsilon_n=\frac{\hbar^2}{2M}{\left(\frac{\pi (n+1)}{d}\right)^2},\quad n=0,1,2,\dots.
\end{equation}
This solution is usually ignored in most of the research on such 
problems \cite{F,GJ,Ng} on account of the interest being confined to the motion 
of the particle in the plane at right angles to the magnetic field.

On the other hand the differential equation satisfied by $\psi(r,\theta)$ 
 describes a two-dimensional particle in the perpendicular homogeneous field 
in the circular gauge. Seting 
$\psi(r,\theta)=\phi(r)e^{il\theta}$ with $l$ an integer, 
one can easily show after some algebraic calculations, that the differential 
equation satisfied by 
\begin{equation}\label{s28}
\phi(r)=\left({\frac{e|B|}{2\hbar c}}\right)^{\abs{l}/2}r^{\abs{l}}e^{- \frac{e|B|}{4\hbar c}r^2}G(\sqrt{\frac{e|B|}{2\hbar c}}r)\end{equation} is
$$\frac{d^2G}{d\xi^2}+\left(\frac{|l|+1}{\xi}-1\right)\frac{dG}{d\xi}+\frac{\lambda-2-2|l|}{4\xi}G=0,$$
where
$\xi=\frac{e|B|}{2\hbar c}r^2$ and $\lambda=\frac{4Mc}{e|B|\hbar}\epsilon_{ml}-2l$. 
This is known as Kummer's differential equation, which has a solution
$$G(\xi)={}_1F_1(\frac{-\lambda+2+2|l|}{4}; |l|+1; \xi)$$
with the eigenvalue condition
$\frac{-\lambda+2+2\abs{l}}{4}=-m$, where $m=0,1,2,\dots$ are the principle quantum numbers  
and $l=0,\pm1,\pm 2,\dots$ are the angular momentum quantum numbers. The eigenvalue condition
 yields the Landau levels
$$\epsilon_{ml}=\frac{e|B|\hbar}{2Mc}(2m+l+\abs{l}+1),$$
and the eigenfunctions become
\begin{equation}\label{s29}
\psi_{m,l}(r,\theta)= N_{ml}r^{\abs{l}}e^{- \frac{e|B|}{4\hbar c}r^2}{}_1F_1(-m;|l|+1; \frac{e|B|}{2\hbar c}r^2)e^{il\theta},
\end{equation}
where $N_{ml}$ is a normalization constant and ${}_1F_1$ is the confluent hypergeometric function defined by
$${}_1F_1(-m;\gamma;z)=\sum\limits_{k=0}^m\frac{(-m)_k}{(\gamma)_k\ k!}z^k.$$
The Pochhammer symbol $(a)_k$ is defined by $(a)_0 = 1$ and $(a)_k = a(a+1)(a+2)\dots(a+k-1)$ for $k = 1,2,3,\dots,$ and may be expressed in terms of the Gamma function by $(a)_k={\Gamma(a+k)/ \Gamma(a),}$ when $a$ is not a negative integer $-m$. In the exceptional cases, $(-m)_k = 0$ if $k > m$ and otherwise $(-m)_k = (-1)^k m!/(m-k)!.$ The normalization constant $N_{ml}$ follows out of the inner product relation
\begin{equation}
\langle \psi_{ml}\mid \psi_{m'l'}\rangle=\int\limits_0^{2\pi}\int\limits_0^\infty\psi_{ ml}(r,\theta)\overline{\psi_{ m'l'}(r,\theta)} rdrd\theta=\delta_{mm^\prime}\delta_{ll^\prime}.
\end{equation}
This yields
$$N_{ml}^{-2}=\left(\frac{2\hbar c}{e|B|}\right)^{\abs{l}+1}\frac{\pi \Gamma(\abs{l}+1)}{(\abs{l}+1)_m}m!,$$
and  by means of the identities
\begin{equation}\label{keyi}
\int_0^\infty r^{2\gamma-1}e^{-s r^2}{}_1F_1(-n;\gamma;s r^2){}_1F_1(-m;\gamma;s r^2)dr=\frac{1}{2}\frac{n!\Gamma(\gamma)}{s^\gamma(\gamma)_n}\delta_{mn}
\end{equation}
and
$$
\int_0^{2\pi}e^{i(l-l^\prime)\theta}d\theta=0 \text{ or } 2\pi \text{ according as } l\neq l^\prime \text{ or } l= l^\prime
$$
we readily conclude that  $\{\psi_{ml}(r,\theta)\}$ is indeed an orthonormal set with respect to 
the measure $rdrd\theta$ where $0\leq \theta<2\pi$. Finally, the spectrum of the 
free Hamiltonian $H_0$ is
\begin{equation}\label{s210}
E(m,l,n)=\frac{e|B|\hbar}{2Mc}(2m+l+\abs{l}+1)+\frac{\hbar^2}{2M}
{\left(\frac{\pi (n+1)}{d}\right)^2},\quad n=0,1,2,\dots.
\end{equation}
We immediately observe that the energy levels $\epsilon_{ml}$ for positive $l$, yield
\begin{equation}\label{s211}
E(m,l,n)=\frac{e|B|\hbar}{2Mc}(2m+2l+1)+\frac{\hbar^2}{2M}{\left(\frac{\pi (n+1)}{d}\right)^2},\quad n=0,1,2,\dots.
\end{equation}
For $l$ negative or zero, we have $\abs{l}+l=0$ which cause the infinite degeneracy of  
Landau levels $\epsilon_{ml}$.  Thereby the spectrum (\ref{s210}) becomes
\begin{equation}\label{s212}
E(m,n)=\frac{e|B|\hbar}{2Mc}(2m+1)+\frac{\hbar^2}{2M}{\left(\frac{\pi (n+1)}{d}\right)^2},\quad n=0,1,2,\dots.
\end{equation}
This particular expression of the spectrum was the starting point of the interesting 
study of Exner and Nemcova \cite{Ex} concerning the spectral properties of a Hamiltonian 
describing the motion of a spinless quantum particle confined to an infinite planar layer with 
hard walls and interacting with a periodic lattice of point perturbations as well as in a homogeneous 
magnetic field perpendicular to the layer.  They remark therein that the spectrum (\ref{s212}) is nondegenerate 
if the ratio of the coefficients $\abs{B}$ and $\pi^2/d^2$ is irrational \cite{Exn}. We shall claim this in the next section.

For simplicity we may assume hereafter that $e=\hbar=2M=c=1$, and hence we  summarize 
the situation as follows. For each $n=0,1,2,\dots$ there is an orthonormal set 
of wavefunctions $\Psi_{mln}(r,\theta,z)\equiv \psi_{ml}(r,\theta)\chi_n(z)$, 
eigensolutions for the Hamiltonian $H_0$, given by
\begin{eqnarray}
\Psi_{mln}(r,\theta,z)&=& \sqrt{\left(\frac{|B|}{2}\right)^{\abs{l}+1}\frac{2(\abs{l}+1)_m}{\pi~d~ m!\Gamma(\abs{l}+1)}}r^{2\abs{l}}e^{-\frac{|B|}{4}r^2}\nonumber\\
&&\hspace{2cm}\times{}_1F_1(-m;|l|+1;\frac{|B|r^2}{2})e^{il\theta}\sin\left(\frac{(n+1)\pi z}{d}\right)\label{s213}
\end{eqnarray}
in the state Hilbert space  $\mathfrak{L}^2(\Sigma)$, which actually is the direct product $\mathfrak{L}^2[0,\infty)\otimes L^2[0,2\pi)\otimes L^2[0,d])$.
\section{Density Argument}
\noindent Making use of the tensor product concept immediately preceeding, we lump the tensor product $ \mathfrak{L}^2[0,\infty)\otimes L^2[0,2\pi)$ of the first two Hilbert spaces  into the Hilbert space $\mathfrak{L}^2([0,\infty)\times [0,2\pi))$, which consists of all complex-valued Lebesgue measurable functions $h$ on $[0,\infty)\times [0,2\pi)$ with
$$\int_0^\infty\int_0^{2\pi} \abs{h(r,\theta)}^2rd\theta dr<\infty.$$

\noindent Let $\mathfrak{L}^2(\Sigma^\prime)\equiv \mathfrak{L}^2([0,\infty)\times [0,2\pi))\otimes L^2[0,d]$, where $$\{\Psi_{ml}(r,\theta)= \sqrt{\left(\frac{|B|}{2}\right)^{\abs{l}+1}
\frac{(\abs{l}+1)_m}{\pi~ m!\Gamma(\abs{l}+1)}}r^{2\abs{l}}e^{-\frac{|B|}{4}r^2}{}_1F_1(-m;|l|+1;\frac{|B|r^2}{2})e^{il\theta}\}$$
with $m=0,1,2,\dots$ and $l=0,\pm 1,\pm 2,\dots$ is an orthonormal system of the Hilbert space $\mathfrak{L}^2([0,\infty)\times [0,2\pi))$ and
$$\bigg\{\sqrt{\frac{2}{d}}~\sin\bigg(\frac{\pi(n+1)z}{d}\bigg): n=0,1,2,\dots\bigg\}$$
is an orthonormal basis of the Hilbert space $L^2[0,d]$. If we can show that $\{\Psi_{ml}: m=0,1,2,\dots, l=0,\pm 1,\pm 2,\dots\}$ is an orthonormal basis of  the Hilbert space $ \mathfrak{L}^2([0,\infty)\times [0,2\pi))$, then $\{\Psi_{mln}: m=0,1,2,\dots, l=0,\pm 1,\pm 2,\dots, n=0,1,2,\dots\}$ becomes an orthonormal basis (\cite{W}, page 52, Theorem 3.12) of the Hilbert space $\mathfrak{L}^2(\Sigma^\prime)$.
\medskip

\begin{theorem}The set $\{\Psi_{ml}: m=0,1,2,\dots, l=0,\pm 1,\pm 2,\dots\}$ is an orthonormal basis of the Hilbert space $\mathfrak{L}^2([0,\infty)\times [0,2\pi))$.
\end{theorem}
\begin{proof}
 Let us assume that it is not. Thus there exist a nontrivial $\Psi\in \mathfrak{L}^2([0,\infty)\times [0,2\pi))$ satisfying
$$\int\limits_0^\infty\int\limits_0^{2\pi} \Psi_{ml}(r,\theta)\overline{\Psi(r,\theta)}rdrd\theta=0\quad\hbox{ for all } m=0,1,2,\dots \hbox{ and } l=0,\pm 1,\pm 2,\dots.$$
Since the linear hull \cite{H}
$$(L.H.) \bigg({}_1F_1(-k;\abs{l}+1;\frac{|B|r^2}{2})(0\leq k\leq m)\bigg)=(L.H.)\bigg(r^{2k}(0\leq k\leq m)\bigg),$$
it follows after taking suitable linear combination of the orthonormal set $\{\Psi_{kl}: 0\leq k\leq n\}$ with $l$ fixed, that
$$\int\limits_0^\infty\int\limits_0^{2\pi} r^{2\abs{l}+2m}e^{-\frac{|B|r^2}{4}}e^{il\theta}\overline{\Psi(r,\theta)}rdrd\theta=\int\limits_0^\infty r^{2\abs{l}+2m+1}e^{-\frac{|B|r^2}{4}}\int\limits_0^{2\pi}e^{il\theta}\overline{\Psi(r,\theta)}d\theta dr=0$$
for all $m=0,1,2,\dots$ and $ l=0,\pm 1,\pm 2,\dots.$ By a further linear combination involving the complex parameter $s$, namely
\begin{equation}\label{sumk}
e_m(sr^2)=\sum_{k=0}^m \frac{(sr^2)^k}{k!},
\end{equation}
we obtain, by means of Lebesgue dominated convergence theorem \cite{R} applied in terms of the following inequality $$\abs{r^{2\abs{l}+1}e_m(-sr^2)e^{-\frac{|B|r^2}{4}}e^{il\theta}}\leq r^{2\abs{l}+1}e^{(\abs{s}-\frac{|B|}{4})r^2}\abs{\Psi(r,\theta)}\in L^1([0,\infty)\times [0,2\pi))$$
for all $m=0,1,2,\dots$ and after taking limit $m\rightarrow \infty$, that the holomorphic function of variable $s$
$$\int\limits_0^\infty\int\limits_0^{2\pi} r^{2\abs{l}+1}e^{-(s+\frac{|B|}{4})r^2}e^{i l\theta}\overline{\Psi(r,\theta)}dr=\int\limits_0^\infty r^{2\abs{l}+1}e^{-(s+\frac{|B|}{4})r^2}\int\limits_0^{2\pi}e^{i l\theta}\overline{\Psi(r,\theta)}d\theta dr=0$$
for all $s$ in the half-plane $\Re(s)>-\frac{|B|}{4}$. We arrive at this conclusion by means of analytic continuation of the fact that the immediate preceeding holomorphic function takes on the value 0 if $\abs{s}< \frac{|B|}{4}$. We make the substitution $r=\sqrt{t}$, and thus achieve
$$ \int\limits_0^\infty t^{\abs{l}}e^{-(s+\frac{|B|}{4})t} \int\limits_0^{2\pi}e^{il\theta}\overline{\Psi(\sqrt{t},\theta)}d\theta dt=0\quad \hbox{ for all }\quad l=0,\pm 1,\pm 2,\dots.$$
Utilizing the uniqueness of Laplace transform \cite{D}, we conclude that
$$\int\limits_0^{2\pi}e^{il\theta}\overline{\Psi(r,\theta)}d\theta =0\quad \hbox{a.e. in }r\hbox{ on }[0,\infty)\hbox{ for } l=0,\pm 1,\pm 2,\dots.$$
In consequence hereof, there exist Lebesgue measurable subsets $E_l$ of $[0,\infty)$, such that their complements in $[0,\infty)$ have one-dimensional Lebesgue measure zero - i.e. $\mu_1((0,\infty)\setminus E_l)=0$ for all $l=0,\pm 1,\pm 2,\dots.$ We define $E=\bigcap\limits_{l=-\infty}^{\infty} E_l$ and note
$$ \int\limits_0^{2\pi}e^{il\theta}\overline{\Psi(r,\theta)}d\theta=0\quad\forall r\in E\; \hbox{ and }\; \mu_1((0,\infty)\setminus E)=0,$$
which follows directly from
$$(0,\infty)\setminus \bigg(\bigcap\limits_{l=-\infty}^\infty E_l\bigg)=\bigcup\limits_{l=-\infty}^\infty \bigg((0,\infty)\setminus E_l\bigg)$$
$$\hbox{ with }\;\;\mu_1((0,\infty)\setminus E)\leq \sum\limits_{l=-\infty}^\infty \mu_1((0,\infty)\setminus E_l)=0.$$
Thus it becomes clear that
$$\int\limits_0^{2\pi}e^{il\theta}\overline{\Psi(r,\theta)}d\theta=0\quad  \forall~r\in E\hbox{ and }\forall~l=0,\pm 1,\pm 2,\dots .$$
Since $\Psi(r,\theta)\in \mathfrak{L}^2([0,\infty)\times [0,2\pi))$, namely
$$\int\limits_0^\infty\int\limits_0^{2\pi} \abs{\Psi(r,\theta)}^2rd\theta dr=\int\limits_0^{2\pi}\int\limits_0^\infty \abs{\Psi(r,\theta)}^2rdrd\theta = \iint\limits_{[0,\infty)\times [0,2\pi)} \abs{\Psi(r,\theta)}^2d\mu(r,\theta)<\infty$$
with $d\mu(r,\theta)=rdrd\theta$, which follows from the Tonelli-Hobson theorem \cite{S}, we may conclude without loss of generality that
$$\int\limits_0^{2\pi}e^{il\theta}\overline{\Psi(r,\theta)}d\theta=0\quad  \forall\quad l=0,\pm 1,\pm 2,\dots \hbox{ and } r\in E\hbox{ with  }\Psi(r,\cdot)\in \mathfrak{L}^2[0,\infty).$$
We consequently have for $r\in E$ with $\Psi(r,\cdot)\in \mathfrak{L}^2[0,\infty)$ that
$$\int\limits_0^{2\pi}\abs{\Psi(r,\theta)}^2d\theta =0
\hbox{ for all $r$ satisfying }
\int_0^{2\pi}\abs{\Psi(r,\theta)}^2d\theta< \infty.$$
Because this holds for almost all $r\in [0,\infty)$, it follows that
$$\int\limits_0^\infty\int\limits_0^{2\pi} \abs{\Psi(r,\theta)}^2rd\theta dr=0,$$
which in turn implies $\Psi$ is a trivial $\mathfrak{L}^2([0,\infty)\times [0,2\pi))$-function.
 Hence $\{\Psi_{ml}: m=0,1,2,\dots, l=0,\pm 1,\pm 2,\dots\}$ is an orthonormal basis of 
$\mathfrak{L}^2([0,\infty)\times [0,2\pi)).$
\end{proof}
Thus $\{\Psi_{mln}: m=0,1,2,\dots, l=0,\pm 1,\pm 2,\dots, n=0,1,2,\dots\}$ is an orthonormal
 basis of $\mathfrak{L}^2(\Sigma^\prime)=\mathfrak{L}^2([0,\infty)\times [0,2\pi))\otimes 
L^2[0,d].$\\
 We shall also consider the case where $\abs{l}+l=0$, where in this case the spectrum 
takes the form
\begin{equation}\label{s51}
E(m,n)=|B|(2m+1)+\left(\frac{\pi(n+1)}{d}\right)^2.
\end{equation}
We fix $l=0$ for the wavefunction $\psi_{mnl}$ of (\ref{s213}). In this case 
$\psi_{mnl}:=\psi_{mn}$ can be written as $\psi_{mn}=\phi_m\otimes\chi_n$ where
$$\phi_{m}(r)=\sqrt{|B|}e^{-\frac{|B|}{4}r^2}{}_1F_1(-m;1;\frac{|B|r^2}{2})\;\;\text{and}
\;\;\chi_n(z)=\sqrt{\frac{2}{d}}\sin{\big(\frac{\pi (n+1)z}{d}\big)}.$$
From (\ref{keyi}) we have
$$\int_{0}^{\infty}e^{-\frac{|B|}{2}r^2}{}_1F_{1}(-n;1;\frac{|B|r^2}{2})
{}_1F_{1}(-m;1;\frac{|B|r^2}{2})rdr=\frac{1}{|B|}\delta_{mn}.$$
Thus $\{\phi_m:m=0,1,2,\dots\}$ is an orthonormal system in the Hilbert space
 $\mathfrak{L}^2[0,\infty).$
\begin{corollary}
When $|B|$ and $\frac{\pi^2}{d^2}$ are irrationally related, the 
spectrum $E(m,n)$ of (\ref{s51}) is nondegenerate and the set of vectors
$$\{\psi_{mn}=\phi_m\otimes\chi_n:m=0,1,2,...;n=0,1,2,...\}$$
forms an orthonormal basis of the Hilbert space $\mathfrak{L}^2[0,\infty)\otimes L^2[0,d]$.
\end{corollary}
\begin{proof}
If we have two pairs $(m,n)$ and $(m',n')$ such that $E(m,n)=E(m',n')$ then
$$\frac{\pi^2}{|B|d^2}=\frac{2(m'-m)}{(n+1)^2-(n'+1)^2}$$
is a rational number. To prove $\B_1=\{\psi_{mn}:m=0,1,2,...;n=0,1,2,...\}$ is an 
orthonormal basis of $\mathfrak{L}^2[0,\infty)\otimes L^2[0,d]$ it is enough to 
show that $\B_2=\{\phi_m:\ m=0,1,2,...\}$ is an orthonormal basis
 of $\mathfrak{L}^2[0,\infty)$. Suppose $\B_2$ is not an orthonormal
 basis of $\mathfrak{L}^2[0,\infty)$, then there exists 
a non-trivial $\phi\in\mathfrak{L}^2[0,\infty)$ such that
$$\int_{0}^{\infty}\phi_m(r)\overline{\phi(r)}rdr=0$$
for all $m=0,1,2,...$. Since 
$$(L.H.)\left({}_1F_1(-k;1;\frac{|B|r^2}{2})~(0\leq k\leq m)\right)=
(L.H.)\left(r^{2k}~(0\leq k\leq m)\right),$$
we have after taking suitable linear combination of the orthonormal 
set $\{\phi_k:0\leq k\leq m\}$ that
$$\int_{0}^{\infty}r^{2m+1}e^{-\frac{|B|r^2}{2}}\overline{\phi(r)}dr
=0;\;\;\text{for all}\;m=0,1,2,....$$
By a further linear combination of (\ref{sumk}) and by means of the Lebesgue dominated
 convergence theorem applied to
$$\abs{e_m(-sr^2)e^{-\frac{|B|r^2}{4}}}\leq e^{|s|-\frac{|B|r^2}{4}}\abs{\phi(r)}
\in L^1[0,\infty);\;\;m=0,1,2,...$$
we obtain
$$\int_{0}^{\infty}e^{-(s+\frac{|B|}{4})r^2}\overline{\phi(r)}rdr=0$$
for all $s$ such that $\mathfrak{R}e(s)>-\frac{|B|}{4}$. By letting $r=\sqrt{t}$ we have
$$ \int_{0}^{\infty}e^{-st}e^{-\frac{|B|t}{4}}\overline{\phi(\sqrt{t})}dt=0$$
for all $s$ such that $\mathfrak{R}e(s)>-\frac{|B|}{4}$. 
 Uniqueness of the Laplace transform yields 
$$e^{-\frac{|B|t}{4}}\overline{\phi(\sqrt{t})}=0\;\; \text{a.e.}\; \text{in}
 \;\;t\;\;\text{ on}\;\; [0,\infty)\;\;\text{ or}\;\; \phi(r)=0\;\;\text{ a.e.}
\;\;\text{in}\;\; r\;\;\text{ on}\;\; [0,\infty)$$ and  consequetly 
$$\int_{0}^{\infty}\abs{\phi(r)}^2rdr=0.$$
Hereby $\phi$ is trivial in $\mathfrak{L}^2([0,\infty))$, which contradict the assumption. 
Thus $\{\phi_m:m=0,1,2,...\}$ is an orthonormal basis of $\mathfrak{L}^2[0,\infty)$. 
\end{proof}
\begin{remark}
Since
$$\bigoplus_{l=-\infty}^{\infty}\mathfrak{L}^2[0,\infty)\otimes\{e^{il\theta}\}
\otimes L^2[0,d]=\mathfrak{L}^2[0,\infty)\otimes L^2[0,2\pi)\otimes L^2[0,d],$$
one can prove for each fixed $l<0$  that the spectrum $E(m,n)$ is non-degenerate 
and the set of vectors $\{\psi_{mnl}:m=0,1,2,...;~n=0,1,2,...;~l~{\text{fixed and}~<0}\}$ 
is an orthonormal basis of the subspace $\mathfrak{L}^2[0,\infty)\otimes\{e^{il\theta}\}
\otimes L^2[0,d]$.
\end{remark}
\section{Gazeau-Klauder coherent states}\label{GKCS}
In this section, we introduce the general features of Gazeau-Klauder CS. Let $H$ be a Hamiltonian with a bounded below discrete spectrum $\{e_m\}_{m=0}^{\infty}$ and it has been adjusted so that $H\geq 0$. Further assume that the
eigenvalues $e_m$ are non-degenerate and arranged in increasing order $\ e_0<e_1<e_2...$. For such a Hamiltonian, a class of CS was
 suggested by Gazeau and Klauder \cite{Gk}, the so-called {\em Gazeau-Klauder
 coherent states} (GKCS for short), as
\begin{equation}\label{s41}
\mid
J,\alpha\rangle=\N(J)^{-1}\sum_{m=0}^{\infty}\frac{J^{m/2}e^{-ie_m\alpha}}{\sqrt{\rho(m)}}
\eta_{m}
\end{equation}
where $J\geq 0,$ $-\infty<\alpha< \infty$,
$\{\eta_m\}_{m=0}^{\infty}$ is the set of eigenfunctions of the
Hamiltonian and $\rho(m)=e_1e_2\dots e_m=e_m!$. 
In order to be GKCS the states (\ref{s41}) need to satisfy the following:
\begin{enumerate}
\item[(a)] For each $J,\alpha$ the state is normalized, i.e.
$1=\langle J,\alpha\mid J,\alpha\rangle = \N(J)^{-2}\sum\limits_{m=0}^{\infty}\frac{J^{m}}
{\rho(m)}
;$
\item[(b)] The set of states $\{\mid J,\alpha\rangle:J\in[0,\infty),\alpha\in(-\infty,\infty)\}$ satisfies a resolution of the identity
\begin{equation}\label{s455}
\lim_{\delta\rightarrow \infty}\frac{1}{2\delta}\int_{-\delta}^{\delta}d\alpha
\int_{0}^{\infty}\lambda(J)dJ\mid J,\alpha\rangle\langle J,\alpha\mid =I
\end{equation}
where $\lambda(J)$ is an appropriate weight function;
\item[(c)] The states are temporally stable, i.e.,
$e^{-iHt}\mid J,\alpha\rangle=\mid J,\alpha+ t\rangle$;
\item[(d)]The states satisfy the action identity, i.e.,
$\langle J,\alpha\mid H\mid J,\alpha\rangle=J$.
\end{enumerate}
The condition (d) requires $e_0=0$. In the case where only the conditions (a)-(c) are satisfied we phrase the resulting CS as ``temporally stable CS". In the case where $e_0\not=0$ one can shift the spectrum backward by $e_0$ and work with the shifted spectrum. \\
The dynamical algebra of the system can be defined as follows: The generalized annihilation, creation and number operators
defined on the state Hilbert space $\mathfrak H$ with respect to the
basis $\{\eta_m\}_{m=0}^{\infty}$ can be given by (see \cite{Ali})
 \begin{eqnarray}
\mathfrak{a} \eta_m &=& \sqrt{e_m} \eta_{m-1}, \quad \text{with}
\quad \mathfrak{a} \eta_0 = 0,\nonumber \\
\mathfrak{a}^\dagger \eta_m &=& \sqrt{e_{m+1}} \eta_{m+1}, \label{s45}\\
\mathfrak{n} \eta_m &=& e_m \eta_m, \quad (\mathfrak{n} =
\mathfrak{a}^\dagger \mathfrak{a})\nonumber
\end{eqnarray}
and the commutators take the form
\begin{eqnarray}\label{s46}
\left[ \mathfrak{a}, \mathfrak{a}^\dagger \right] \eta_m
&=& (e_{m+1}-e_m) \eta_m, \nonumber \\
\left[ \mathfrak{n}, \mathfrak{a}^\dagger \right] \eta_m &=&
(e_{m+1}-e_m) \mathfrak{a}^\dagger \eta_m, \\
\left[ \mathfrak{n}, \mathfrak{a} \right] \eta_m &=& (e_{m-1} -
e_m) \mathfrak{a} \eta_m. \nonumber
 \end{eqnarray}
The algebra generated by the operators $\{\an,\an^{\dagger},\n\}$ and its deformations
(up to isomorphisms) serve as a dynamical algebra of the Hamiltonian. \\
In \cite{Ng} the definition (\ref{s41}) was generalized to multi-dimensions as
\begin{equation}\label{s47}
\mid\mathbf{J},{\mathbf{b}}\rangle=\N(\mathbf{J})^{-1}\sum_{\{n_1,...,n_r\}}\frac{\mathbf{J}^{\mathbf{n}/2}}{\sqrt{\rho(\mathbf{n})}}e^{-i\mathbf{b}\cdot e(\mathbf{n})}\mid\mathbf{n}\rangle
\end{equation}
where the sum runs over all possible values of the variables $n_j$, $\N$ is a normalization factor and $\rho(\mathbf{n})$ is an arbitrary positive function of all the indices. Further, $\mathbf{J}^{\mathbf{n}/2}=\prod_{j=1}^{r}J_j^{n_j/2}$, $\mathbf{b}\cdot e(\mathbf{n})=\alpha_1e_1(\mathbf{n})+...+\alpha_re_r(\mathbf{n})$ and $\mid \mathbf{n}\rangle=\mid n_1\rangle\otimes...\otimes\mid n_r\rangle$ where $\{\mid n_j\rangle\}$ forms an orthonormal basis for an appropriate Hilbert space $\HI_j$. Using (\ref{s47}) GKCS for the rth degree of freedom is defined as
\begin{equation}\label{s48}
\mid n_1,...,n_{r-1},J_r,\alpha_r\rangle=\N_r(J_r)^{-1}\sum_{n_r}\frac{J_r^{n_r/2}}{\sqrt{\rho_r}}e^{-i\alpha_re_r(\mathbf{n})}\mid \mathbf{n}\rangle
\end{equation}
where the normalization factor $\N_r$ and the function $\rho_r$ may depend on the
 other indices. In addition to the normalization condition, when $n_1,...,n_{r-1}$ are fixed, 
the states (\ref{s48}) should satisfy a resolution of the identity on the subspace obtained by
 fixing $n_1,n_2,..n_{r-1}$:
\begin{equation}\label{s49}
\int \mid n_1,...,n_{r-1},J_r,\alpha_r\rangle\langle n_1,...,n_{r-1},J_r,
\alpha_r\mid d\mu(J_r,\alpha_r)=I_{n_1,n_2,..n_{r-1}}.
\end{equation}
For the multi-dimensional case, if one takes $\rho_j(\mathbf{n})=\rho_j(n_1,n_2,...,n_j)$ and $e_j(\mathbf{n})=e_j(n_1,n_2,...,n_j)$ we can associate multiple degrees of freedom:
\begin{equation}\label{s410}
\mid\mathbf{J},\mathbf{b}\rangle=\N_1^{-1}\sum_{n_1}\frac{J_1^{n_1/2}}{\sqrt{\rho_1}}e^{-i\alpha_1e_1}\N_2^{-1}\sum_{n_2}\frac{J_2^{n_2/2}}{\sqrt{\rho_2}}e^{-i\alpha_2e_2}...\N_r^{-1}\sum_{n_r}\frac{J_r^{n_r/2}}{\sqrt{\rho_r}}e^{-i\alpha_re_r}\mid\mathbf{n}\rangle
\end{equation}
where $\N_j=\N_j(J_j,...,J_r;n_1,...,n_{j-1})$. If $\rho_j$ and $e_j$ are independent of $n_k,$ $k<j$ then the states (\ref{s410}) may give us simple tensor product of states. For the states (\ref{s410}) a resolution of the identity takes the following form:
\begin{eqnarray*}
&&\lim_{\delta\rightarrow\infty}\frac{1}{2\delta}\int_{-\delta}^{\delta}d\alpha_1\int_{0}^{\infty}\lambda_1(J_1)\lim_{\delta\rightarrow\infty}\frac{1}{2\delta}\int_{-\delta}^{\delta}d\alpha_2\int_{0}^{\infty}\lambda_1(J_1,J_2,n_1)
...\lim_{\delta\rightarrow\infty}\frac{1}{2\delta}\int_{-\delta}^{\delta}d\alpha_r\\
&&\times\int_{0}^{\infty}\lambda_r(J_1,...,J_{r},n_1,...,n_{r-1})
\mid n_1,...,n_{r},\mathbf{J},\mathbf{b}\rangle\langle n_1,...,n_{r},\mathbf{J},\mathbf{b}\mid 
dJ_1...dJ_r\\
&&=I_{\HI_1}\otimes...\otimes I_{\HI_r}
\end{eqnarray*}
where $\lambda_j,~ j=1,2,...,r$ are positive weight functions. For the multi-dimensional case, 
the temporal stability and the action identity can also be added. We will discuss 
these issues through the problem of this paper in Section 6.\\
In the following sections, when $l=0$ we derive temporally stable CS for the Hamiltonian
 $H_0$ with the spectrum $E(m,n)$ on the subspace $\mathfrak{L}^2[0,\infty)\otimes\frac{1}{\sqrt{2\pi}}\otimes L^2[0,d]$, which is indeed a subspace of $\mathfrak{L}^2(\Sigma)$. However, $\mathfrak{L}^2[0,\infty)\otimes\frac{1}{\sqrt{2\pi}}\otimes L^2[0,d]$ is isomorphic (in the Hilbert space sense) to $\frac{1}{\sqrt{2\pi}}\otimes\mathfrak{L}^2[0,\infty)\otimes L^2[0,d]$ as subspaces of $\mathfrak{L}^2[0,\infty)\otimes L^2[0,2\pi]\otimes L^2[0,d]$ and $L^2[0,2\pi]\otimes \mathfrak{L}^2[0,\infty)\otimes L^2[0,d]$ respectively. Nevertheless, the subspace $\frac{1}{\sqrt{2\pi}}\otimes\mathfrak{L}^2[0,\infty)\otimes L^2[0,d]$  is (Hilbert space) isomorphic to $\mathfrak{L}^2[0,\infty)\otimes L^2[0,d]$ and hence, we may consider the Hilbert space $\mathfrak{L}^2[0,\infty)\otimes L^2[0,d]$ instead of the subspace $\mathfrak{L}^2[0,\infty)\otimes\frac{1}{\sqrt{2\pi}}\otimes L^2[0,d]$ to which it is isomorphic. Therefore, the action of the Hamiltonian $H_0$ on $\mathfrak{L}^2[0,\infty)\otimes\frac{1}{\sqrt{2\pi}}\otimes L^2[0,d]$ carries to the Hilbert space $\mathfrak{L}^2[0,\infty)\otimes L^2[0,d]$. Hereafter we refer the Hilbert space $\mathfrak{L}^2[0,\infty)\otimes L^2[0,d]$ as the state Hilbert space of the Hamiltonian $H_0$ for $l=0$.\\
The GKCS studied in \cite{F} for the Landau levels may be regarded as a set of GKCS constructed in the absence of $n$ and $l$ from the spectrum $E(m,l,n)$ of (\ref{s210}). The GK-like CS (in the terminology of \cite{Ng}) studied in \cite{Ng} can be taken as a class of CS for the spectrum $E(m,l,n)$ in the absence of $n$.
\section{CS with one degree of freedom}\label{1DCS}
We introduce two classes of temporally stable CS with the form (\ref{s48}) 
for the spectrum (\ref{s51}), by first fixing $n$ followed by another class 
where $m$ is fixed. In both cases the orthonormal basis is denoted by the same 
symbol $\psi_{m,n}$. However, it should be clear that in each case the other index 
is fixed and the vectors $\psi_{m,n}$ belong to the corresponding subspace of the state
 Hilbert space $\mathfrak{L}^2[0,\infty)\otimes L^2[0,d]$. For sake of simplicity $|B|:=B$. 
\subsection{When $n$ is fixed}\label{NFIX}
Here we discuss a class of temporally stable CS for the first degree 
of freedom (the freedom through $m$). Since the energy spectrum of the
 Landau problem is $E_m=B(2m+1)$, the following set of CS can also be considered 
as a set of CS with a forward shift of the Landau levels. Further, the construction also serves as a preparatory step of the formation of CS for two degrees of freedom. Let
$$\rho(m)=E(1,n)E(2,n)...E(m,n)$$
where $E(m,n)$ is given by (\ref{s51}).
We have
$$\rho(m)=\prod_{k=1}^{m}\left(B(2k+1)+\left(\frac{\pi (n+1)}{d}\right)^2\right)
=(2B)^m(\gamma)_m$$
where
$$\gamma=1+\frac{Bd^2+\pi^2(n+1)^2}{2Bd^2}.$$
Let us study the following class of vectors
\begin{equation}\label{s52}
\mid J,\alpha,n\rangle=\N(J,n)^{-1}\sum_{m=0}^{\infty}\frac{J^{m/2}e^{-iE(m,n)\alpha}}{\sqrt{\rho(m)}}\psi_{m,n}.
\end{equation}
The normalization condition $\langle J,\alpha,n\mid J,\alpha,n\rangle=1$ yields
\begin{equation}\label{s53}
\N(J,n)^2=\sum_{m=0}^{\infty}\frac{J^m}{2^mB^m(\gamma)_m}={}_1F_1(1;\gamma;\frac{J}{2B})>0,
\end{equation}
which converges for all $J>0$. For a resolution of the identity, let $-\infty<\alpha<\infty$ and set a measure
$$d\mu(J,\alpha)=d\nu(J)d\alpha=\frac{1}{2^{\gamma}B^\gamma\Gamma(\gamma)}
 {}_1F_1(1;\gamma;\frac{J}{2B})e^{-\frac{J}{2B}}J^{\gamma-1}dJd\alpha.$$
The knowledge of the equations (\ref{s455}) and (\ref{s49}) leads to
\begin{eqnarray*}
&&\int_{0}^{\infty}\int\mid J,\alpha,n\rangle\langle J,\alpha,n\mid d\mu(J,\alpha)\\
&&=\sum_{m=0}^{\infty}\sum_{l=0}^{\infty}\frac{\mid\psi_{m,n}\rangle\langle\psi_{l,n}\mid}
{\sqrt{\rho(m)\rho(l)}}\int_{0}^{\infty}\int\frac{J^{m/2+l/2}}{\N(J,n)^2}e^{i(E(m,n)-E(l,n))
\alpha}d\nu(J)d\alpha\\
&&=\sum_{m=0}^{\infty}\frac{\mid\psi_{m,n}\rangle\langle\psi_{m,n}\mid}{2^{m+\gamma}B^{m+\gamma}(\gamma)_m\Gamma(\gamma)}
\int_{0}^{\infty}J^{m+\gamma-1}e^{-\frac{J}{2B}}dJ\\
&&=\sum_{m=0}^{\infty}\mid\psi_{m,n}\rangle\langle\psi_{m,n}\mid=I_{n},
\end{eqnarray*}
where we employed the identity
\begin{equation}\label{s54}
\int_{0}^{\infty}e^{-ax}x^{s-1}dx=a^{-s}\Gamma(s)
\end{equation}
with $s=m+\gamma$ and $a=\frac{1}{2B}$. For the temporal stability, since for fixed $n$
$$H_0\psi_{m,n}=E({m,n})\psi_{m,n},\;\;\;\;m=0,1,2,...$$
and
$$e^{-iE(m,n)\alpha}e^{-iH_0t}\psi_{m,n}=e^{-iE(m,n)\alpha}e^{-iE(m,n)t}\psi_{m,n}
=e^{-iE(m,n)(\alpha+t)}\psi_{m,n}$$
we have
\begin{equation}\label{s55}
e^{-iH_0t}\mid J,\alpha,n\rangle=\mid J,\alpha+t,n\rangle.
\end{equation}
Thus the states $\mid J,\alpha,n\rangle$ form a set of temporally stable CS.
 Since $E(0,n)\not=0$ the action identity cannot be obtained. The overlap of two states takes the form
\begin{eqnarray*}
\langle J,\alpha,n\mid J',\alpha',n\rangle&=&\frac{e^{-i(\alpha-\alpha^\prime)(B+\frac{\pi^2(n+1)^2}{d^2})}}{\sqrt{{}_1F_1(1;\gamma;\frac{J}{2B}){}_1F_1(1;\gamma;\frac{J'}{2B})}}{}_1F_1(1;\gamma;\frac{JJ^\prime e^{-2iB(\alpha-\alpha^\prime)}}{2B}).
\end{eqnarray*}
If $\alpha=\alpha'$ we get
\begin{equation}
\langle J,\alpha,n\mid J',\alpha,n\rangle=
\frac{{}_1F_1(1;\gamma;\frac{\sqrt{JJ'}}{2B})}{\sqrt{{}_1F_1(1;\gamma;\frac{J}{2B})
{}_1F_1(1;\gamma;\frac{J'}{2B})}}.\nonumber
\end{equation}

\begin{remark}
$\bullet$~In (\ref{s52}) instead of taking $\rho(m)=E(1,n)\dots E(m,n)$ if we take
$\rho(m)=e(1,m)\dots e(m,n)$ with $e(m,n)=E(m,n)-E(0,n)$ then we can have
$0=e(0,n)<e(1,n)...$ and thereby we can have a set of GKCS. In this case,
$\rho(m)=2^mB^mm!$, $\N(J)^2=e^{J/(2B)}$ and a resolution of the identity is
 obtained with the measure
 $d\mu(J,\alpha)=\frac{1}{2B\N^2(J)}e^{-J/(2B)}d\alpha dJ$. The
 temporal stability and the action identity follow straightforwardly.\\
$\bullet$~The spectrum of the isotonic oscillator
$$H=-\frac{d^2}{dx^2}+x^2+\frac{A}{x^2}\;\;\;(A\geq 0)$$
is $\mathfrak{e}_m=2(2m+\gamma)$ where $\gamma=1+\frac{1}{2}\sqrt{1+4A}$. Since this spectrum is nondegenerate and the eigenfunctions form an orthonormal basis of the Hilbert space $L^2[0,\infty)$ \cite{HNA},  
when $B=2$ and $\psi_{mn}$ is replaced by the wavefunctions of $H$, the set of CS 
given in (\ref{s52}) can also be considered as a set of temporally stable CS for $H$ with
 a forward shift of the spectrum. 
\end{remark}

\subsection{When $m$ is fixed}\label{MFIX}
 We discuss a class of temporally stable CS for the 2nd degree of 
freedom obtained through $n$ by fixing $m$. That is, 
the following class of CS can be considered as a class of CS constructed with 
the effective part of the spectrum due to the infinite layer. The other aim of this 
subsection is 
to facilitate the calculations of the following sections. For fixed $m$ let
$$\rho(n)=E(m,1)E(m,2)...E(m,n).$$
Thereby
$$\rho(n)=\prod_{k=1}^{n}\left(B(2m+1)+\left(\frac{\pi (k+1)}{d}\right)^2\right)
=\left(\frac{\pi}{d}\right)^{2n}(\beta)_n(\overline{\beta})_n$$
where
$$\beta=2+\frac{id}{\pi}\sqrt{B(2m+1)}$$
and $\overline{\beta}$ is the complex conjugate of $\beta$.
Note that, the product $(\beta)_n(\overline{\beta})_n$ is a real positive number.
Consider the set of vectors
\begin{equation}\label{s56}
\mid J,\alpha,m\rangle=\N(J,m)^{-1}\sum_{n=0}^{\infty}\frac{J^{n/2}}{\sqrt{\rho(n)}}e^{-iE(m,n)\alpha}\psi_{m,n}.
\end{equation}
The normalization factor $\N(J,m)$ is obtained, by demanding 
$\langle J,\alpha,m\mid J,\alpha,m\rangle=1$, in the following form.
\begin{equation}\label{s57}
\N(J,m)^2=\sum_{n=0}^{\infty}\frac{J^n}{\rho(n)}=
\sum_{n=0}^{\infty}\frac{\left(\frac{d^2J}{\pi^2}\right)^n}{(\beta)_n(\overline{\beta})_n}
={}_1F_2\left(1;\overline{\beta},\beta;\frac{d^2J}{\pi^2}\right),
\end{equation}
which is a real positive function and defined for all $J\geq 0$. For $J>0$ and $-\infty<\alpha<\infty$ set
$$d\mu(J,\alpha)=\N(J,m)^2\lambda(J)dJd\alpha.$$
For a resolution of the identity, we have
\begin{eqnarray*}
\int_{0}^{\infty}\int\mid J,\alpha,m\rangle\langle J,\alpha,m\mid d\mu(J,\alpha)
&=&\sum_{n=0}^{\infty}\frac{\mid\psi_{m,n}\rangle\langle\psi_{m,n}\mid}{\rho(n)}
\int_{0}^{\infty}J^n\lambda(J)dJ\\
&=&\sum_{n=0}^{\infty}\mid\psi_{m,n}\rangle\langle\psi_{m,n}\mid=I_{m}
\end{eqnarray*}\label{s58}
if there is a density $\lambda(J)$ to satisfy
\begin{equation}\label{s59}
\int_{0}^{\infty}J^n\lambda(J)dJ=\rho(n)=\left(\frac{\pi}{d}\right)^{2n}
(\beta)_n(\overline{\beta})_n.
\end{equation}
Since
\begin{equation}\label{s510}
\int_{0}^{\infty}2K_{2\eta}(2\sqrt{x})x^{s-1}dx=\Gamma(s-\eta)\Gamma(s+\eta)
\end{equation}
the density
\begin{equation}\label{s511}
\lambda(J)=\frac{2d^2}{\pi^2\Gamma({\beta})\Gamma(\overline\beta)}K_{\beta-\overline{\beta}}\left(\frac{2d}{\pi}\sqrt{J}\right)
\end{equation}
satisfies (\ref{s59}), where $K$ is the modified Bessel function of the third kind of imaginary order \cite{key10} and may be regarded as the kernel of the Kontorovich-Lebedev transform \cite{Nay} in the light of (\ref{s510}). The temporal stability follows similar to the previous case. Thus we have a set of temporally stable CS without the action identity. As in the previous case, when $\alpha=\alpha'$ the overlap of two states takes the form
\begin{eqnarray*}
\langle J,\alpha,m\mid J',\alpha,m\rangle&=&\frac{{}_1F_2(1;\overline{\beta},\beta;\frac{d^2\sqrt{JJ'}}{\pi^2})}{\sqrt{{}_1F_2(1;\overline{\beta},\beta;\frac{d^2J}{\pi^2}){}_1F_2(1;\overline{\beta},\beta;\frac{d^2J'}{\pi^2})}}.
\end{eqnarray*}
\begin{remark}
$\bullet$~Let $E_n=E(m,n)-E(m,0)$. In (\ref{s56}) if we replace the $\rho(n)=E(m,1)...E(m,n)$ by $\rho(n)=E_1...E_n=n!(n+2)!/2$ we can have the action identity and thereby a class of GKCS. In this case the normalization factor takes the form $\N(J)^2=J/[2I_2(2\sqrt{J})]$ and a resolution of the identity can be obtained with the measure $d\mu(J,\alpha)=\N(J)^2\lambda(J)d\alpha dJ$ where $$\lambda(J)=\frac{1}{2}G_{0,2}^{2,0}\left( J~\vert~
\begin{array}{c}
- \\ 
2,0
\end{array}
\right),$$
which is given in terms of the MeijerG-function (see \cite{M}, pp. 303, formula (37)).\\
$\bullet$~A class of GKCS for the infinite well potential with the spectrum $\mathbf{e}_n=n(n+2)$ is given in \cite{Tem}. When $d=\pi$ the above class of CS can be considered as a class of temporally stable CS for the infinite well with a forward shift of the spectrum. In this case the state Hilbert space has to be replaced by the Hilbert space of the infinite well.
\end{remark} 
\section{CS with two degrees of freedom}
In this section we present two different classes of CS with two degrees of freedom in the form (\ref{s410}). In the first case, we present a class of CS as a tensor product of two classes of states by setting $\rho_1,\rho_2,e_1$ and $e_2$ independent. In the second case, within the multiple sum, by letting one sum depends on the other through $\rho_1,\rho_2,e_1$ and $e_2$, we present a class of CS where the resulting CS cannot be considered as a tensor product of two states. Further, both classes are considered as temporally stable CS for the Hamiltonian $H_0$ with the spectrum $E(m,n)$.
\subsection{When summations are independent}\label{PTEN}
Let $e_m=B(2m+1)$, $\epsilon_n=[\pi(n+1)/d]^2$,
$\rho_1(m)=e_1e_2...e_m=e_m!,\;\;\text{and}\;\;\rho_2(n)=\epsilon_1\epsilon_2...\epsilon_n=\epsilon_n!.$
Thus
\begin{eqnarray*}
\rho_1(m)&=&\prod_{k=1}^{m}[B(2k+1)]=2^mB^m\left(\frac{3}{2}\right)_m,\\
\rho_2(n)&=&\prod_{j=1}^{n}\left(\frac{\pi (j+1)}{d}\right)^2=\left(\frac{\pi}{d}\right)^{2n}(2)_n(2)_n.
\end{eqnarray*}
The set of vectors under consideration is as follows:
\begin{eqnarray}\label{s61}&&
\mid J_1,J_2,\alpha_1,\alpha_2\rangle=\N_1(J_1)^{-1}\N_2(J_2)^{-1}\bigg[\sum_{m=0}^{\infty}\frac{J_{1}^{m/2}}{\sqrt{\rho_1(m)}}e^{-ie_m\alpha_1}\label{GKCS-1}\\
&~&\hspace{3.5cm}\times~\sum_{n=0}^{\infty}\frac{J_{2}^{n/2}}{\sqrt{\rho_2(n)}}e^{-i\epsilon_n\alpha_2}\phi_{m}\otimes\chi_n\bigg].\nonumber
\end{eqnarray}
Since
$$\langle J_1,J_2,\alpha_1,\alpha_2\mid J_1,J_2,\alpha_1,\alpha_2\rangle=\N_1(J_1)^{-2}\sum_{m=0}^{\infty}\frac{J_{1}^{m}}{\rho_1(m)}\N_2(J_2)^{-2}\sum_{n=0}^{\infty}\frac{J_{2}^{n}}{\rho_2(n)}$$
the normalization requirement $\langle J_1,J_2,\alpha_1,\alpha_2\mid J_1,J_2,\alpha_1,\alpha_2\rangle=1$ yields
$$\N_2(J_2)^2=\sum_{n=0}^{\infty}\frac{J_{2}^{n}}{\rho_2(n)}=\sum_{n=0}^{\infty}\frac{1}{(2)_n ~(2)_n}\left(\frac{d^2J_2}{\pi^{2}}\right)^n={}_1F_2(1;2,2;\frac{d^2J_2}{\pi^2})$$
and
\begin{eqnarray*}
\N_1(J_1)^2=
\sum_{m=0}^{\infty}\frac{J_{1}^{m}}{2^mB^m\left(\frac{3}{2}\right)_m}={}_1F_1(1;\frac{3}{2};\frac{J_1}{2B}).
\end{eqnarray*}
For $J_1,J_2\in(0,\infty)$ and $-\infty<\alpha_1,\alpha_2<\infty$, let us assume that the measure
\begin{equation}\label{s6m}
d\mu(J_1,J_2,\alpha_1,\alpha_2)=\N_1(J_1)^2\N_2(J_2)^2\lambda_1(J_1)\lambda_2(J_2)dJ_1dJ_2d\alpha_1d\alpha_2.
\end{equation}
The weight functions $\lambda_1(J_1)$ and $\lambda_2(J_2)$ will be chosen to satisfy a resolution of the identity. In this case, we have
\begin{eqnarray*}
&&\int_{0}^{\infty}\int_{0}^{\infty}\int\int\mid J_1,J_2,\alpha_1,\alpha_2\rangle\langle J_1,J_2,\alpha_1,\alpha_2\mid d\mu(J_1,J_2,\alpha_1,\alpha_2)\\
&&=\sum_{m=0}^{\infty}\frac{\mid\phi_m\rangle\langle\phi_m\mid}{\rho_1(m)}\int_{0}^{\infty}J_{1}^{m}\lambda_1(J_1)dJ_1\otimes\sum_{n=0}^{\infty}\frac{\mid\chi_n\rangle\langle\chi_n\mid}{\rho_2(n)}\int_{0}^{\infty}J_{2}^{n}\lambda_2(J_2)dJ_2\\
&&=\sum_{m=0}^{\infty}\mid\phi_m\rangle\langle\phi_m\mid\otimes\sum_{n=0}^{\infty}\mid\chi_n\rangle\langle\chi_n\mid
=I_{\mathfrak{L}^2[0,\infty)}\otimes I_{L^2[0,d]}
\end{eqnarray*}
under the assumption that the densities $\lambda_1(J_1)$ and $\lambda_2(J_2)$ are such that
\begin{eqnarray}
\int_{0}^{\infty}J_{1}^{m}\lambda_1(J_1)dJ_1&=&\rho_1(m)=2^mB^m\left(\frac{3}{2}\right)_m\label{s62}\;\;\text{and}\\
\int_{0}^{\infty}J_{2}^{n}\lambda_2(J_2)dJ_2&=&\rho_2(n)=\left(\frac{\pi}{d}\right)^{2n}(2)_n(2)_n.
\label{s63}
\end{eqnarray}
The density
$$\lambda_1(J_1)=\sqrt{\frac{J}{2\pi B^3}}e^{-\frac{J}{2B}},$$
satisfies (\ref{s62}) and the density
$$\lambda_2(J_2)=\frac{2d^4}{\pi^4J_2}K_{0}\left(\frac{2d\sqrt{J_2}}{\pi}\right),$$
where $K_0$ is the modified Bessel function of order $0$, will prove (\ref{s63}). Since $\psi_{m,n}=\phi_m\otimes\chi_n$ and
$H_0\psi_{m,n}=E(m,n)\psi_{m,n}$ we have
$$H_0(\phi_m\otimes\chi_n)=(e_m+\epsilon_n)\phi_m\otimes\chi_n.$$
Therefore, we have
$$e^{-iH_0t}\phi_m\otimes\chi_n=e^{-i(e_m+\epsilon_n)t}\phi_m\otimes\chi_n$$
and thereby
$$e^{-iH_0t}\mid J_1,J_2,\alpha_1,\alpha_2\rangle=\mid J_1,J_2,\alpha_1+t,\alpha_2+t\rangle.$$
Thus the states $\mid J_1,J_2,\alpha_1,\alpha_2\rangle$ are temporally stable.
\begin{remark}\label{re6}
 $\bullet$~Since $H_0\phi_m\otimes\chi_n=(e_m+\epsilon_n)\phi_m\otimes\chi_n,$ even under the assumption $e_0=\epsilon_0=0$ (i.e, even if we shift the spectrum backward), we cannot have the action identity. Therefore, we only have a set of temporally stable CS.\\
$\bullet$~If we shift $e_m$ and $\epsilon_n$ backward by $e_0$ and $\epsilon_0$ we get $\widetilde{e}_m=e_m-e_0=2Bm$ and $\widetilde{\epsilon}_n=\epsilon_n-\epsilon_0=\pi^2 n(n+2)/d^2$ and thereby $\widetilde{\rho}_1(m)=\widetilde{e}_1...\widetilde{e}_m=2^mB^mm!$ and $\widetilde{\rho}_2(n)=\widetilde{\epsilon}_n...\widetilde{\epsilon}_n=\pi^{2n}n!(n+2)!/(2d^{2n})$. In (\ref{s61}) when we replace $\rho_1(m),\rho_2(n),e_m$ and $\epsilon_n$ by $\widetilde{\rho}_1(m),\widetilde{\rho}_2(n),\widetilde{e}_m$ and $\widetilde{\epsilon}_n$ we get
$$\widetilde{\N}_2(J_2)^2={}_0F_1(-;3;\frac{J_2d^2}{\pi^2})=\frac{2\pi^2}{J_2d^2}I_2\left(\frac{2d\sqrt{J_2}}{\pi}\right)\;\;\text{and}\;\;\widetilde{\N}_1(J_1)^2=e^{\frac{J_1}{2B}}.$$
In this case, a resolution of the identity is obtained with the measure
$$d\mu(J_1,J_2,\alpha_1,\alpha_2)=\widetilde{\N}_1(J_1)^2\widetilde{\N}_2(J_2)^2\widetilde{\lambda}_1(J_1)\widetilde{\lambda}_2(J_2)dJ_1dJ_2d\alpha_1d\alpha_2,$$
where
$$\widetilde{\lambda}_1(J_1)=\frac{1}{2B}e^{-J_1/(2B)}\;\;\text{and}\;\;\widetilde{\lambda}_2(J_2)=\frac{\pi^2}{2d^2}G_{0,2}^{2,0}\left( \frac{J_2\pi^2}{d^2}~\vert~
\begin{array}{c}
- \\ 
2,0
\end{array}
\right).$$
The temporal stability follows easily.
\end{remark}
\subsection{When summations depend one on the other}\label{SUMDEP}
For fixed $m$ let
$$\rho_1(m,n)=E(m,1)E(m,2)...E(m,n).$$
From section \ref{MFIX} we have
$$\rho_1(m,n)=\left(\frac{\pi}{d}\right)^{2n}(\beta)_n(\overline{\beta})_n$$
where $\beta$ and $\overline{\beta}$ are as in section \ref{MFIX}. Let
$$\rho_2(m)=e_1...e_m=2^mB^m\left(\frac{3}{2}\right)_m.$$
Consider the following set of vectors
\begin{eqnarray}
&&\mid J_1,J_2,\alpha_1,\alpha_2\rangle=\N_1(J_1,J_2)^{-1}\sum_{m=0}^{\infty}\frac{J_{1}^{m/2}}{\sqrt{\rho_2(m)}}e^{-ie_m\alpha_1}\N_2(J_2,m)^{-1}\nonumber\\
&&\hspace{2cm}\times\sum_{n=0}^{\infty}\frac{J_{2}^{n/2}}{\sqrt{\rho_1(m,n)}}e^{-i\epsilon_n\alpha_2}\phi_{m}\otimes\chi_n.\label{s64}
\end{eqnarray}
In order to obtain the normalization factor let us compute the norm of the vector $\mid J_1,J_2,\alpha_1,\alpha_2\rangle$.
\begin{eqnarray*}
&&\langle  J_1,J_2,\alpha_1,\alpha_2 \mid J_1,J_2,\alpha_1,\alpha_2\rangle=\N_1(J_1,J_2)^{-2}\sum_{m=0}^{\infty}\frac{J_{1}^{m}}{\rho_{2}(m)}\N_2(J_2,m)^{-2}\\
&&\hspace{5cm}\times\sum_{n=0}^{\infty}\frac{J_{2}^{n}}{\rho_1(m,n)}=1
\end{eqnarray*}
if
\begin{eqnarray}
\N_2(J_2,m)^2&=&\sum_{n=0}^{\infty}\frac{J_{2}^{n}}{\rho_1(m,n)}\;\;\;\;\text{and}\label{s65}\\
\N_1(J_1,J_2)^2&=&\sum_{m=0}^{\infty}\frac{J_{1}^{m}}{\rho_{2}(m)\N_2(J_2,m)^2}.\label{s66}
\end{eqnarray}
By (\ref{s57}) we have
$$\N_2(J_2,m)^2={}_1F_{2}\left(1;\beta,\overline{\beta};\frac{d^2J_2}{\pi^2}\right)\geq 1 \quad\quad\quad \forall\quad J_2\in (0,\infty).$$
Thus, we have
\begin{eqnarray*}
\N_1(J_1,J_2)^2&=&\sum_{m=0}^{\infty}\frac{J_{1}^{m}}{\rho_{2}(m){}_1F_{2}\left(1;\overline{\beta},\beta;\frac{d^2J_2}{\pi^2}\right)}\\
&\leq &\sum_{m=0}^{\infty}\frac{J_{1}^{m}}{\rho_{2}(m)} =
\sum_{m=0}^{\infty}\frac{\left(\frac{J_{1}}{2B}\right)^{m}}{\left(\frac{3}{2}\right)_m}={}_1F_1(1;\frac{3}{2};\frac{J_1}{2B}),
\end{eqnarray*}
which converges for all $J_1\geq 0$. For $J_1,J_2\in[0,\infty)$ and $-\infty<\alpha_1,\alpha_2<\infty$ we have
\begin{eqnarray*}
&&\int\limits_{0}^{\infty}\int\limits_{0}^{\infty}\int\int \mid J_1,J_2,\alpha_1,\alpha_2\rangle\langle J_1,J_2,\alpha_1,\alpha_2\mid \lambda_1(J_1)\lambda_2(J_2,m)dJ_1dJ_2d\alpha_1d\alpha_2\\
&&=\sum_{m=0}^{\infty}\sum_{n=0}^{\infty}\frac{\mid\phi_m\rangle\langle\phi_m\mid\otimes\mid\chi_n\rangle\langle\chi_n\mid}{\rho_2(m)\rho_1(m,n)}\int_{0}^{\infty}\frac{J_{1}^{m}}{\N_{1}(J_1,J_2)^2}\lambda_1(J_1)dJ_1\\
&&\hspace{5cm}\times\int_{0}^{\infty}\frac{J_{2}^{n}}{\N_{2}(J_2,m)^2}\lambda_2(J_2,m)dJ_2\\
&&=\sum_{m=0}^{\infty}\sum_{n=0}^{\infty}\mid\phi_m\rangle\langle\phi_m\mid\otimes\mid\chi_n\rangle\langle\chi_n\mid=I_{\mathfrak{L}^2[0,\infty)}\otimes I_{L^2[0,d]}
\end{eqnarray*}
if there are densities $\lambda_1(J_1)$ and $\lambda_2(J_2,m)$ such that
\begin{equation}\label{s67}
\int_{0}^{\infty}\frac{J_{1}^{m}}{\N_{1}(J_1,J_2)^2}\lambda_1(J_1)dJ_1\int_{0}^{\infty}\frac{J_{2}^{n}}{\N_{2}(J_2,m)^2}\lambda_2(J_2,m)dJ_2=\rho_2(m)\rho_1(m,n).
\end{equation}
Let
$$\lambda_2(J_2)=\N_{2}(J_2,m)^2\Lambda_2(J_2,m)\;\;\;\text{and}\;\;\;\lambda_1(J_1)=\N_{1}(J_1,J_2)^2\Lambda_1(J_1).$$
Then (\ref{s67}) reduces to
\begin{equation}\label{s68}
\int_{0}^{\infty}J_{1}^{m}\Lambda_1(J_1)dJ_1\int_{0}^{\infty}J_{2}^{n}\Lambda_2(J_2,m)dJ_2=\rho_2(m)\rho_1(m,n).
\end{equation}
If we combine (\ref{s62}) and (\ref{s63}) we can have (\ref{s68}). Thus we have a resolution of the identity. By the same argument of  subsection \ref{PTEN} we have
$$e^{-iH_0t}\mid J_1,J_2,\alpha_1,\alpha_2\rangle=\mid J_1,J_2,\alpha_1+t,\alpha_2+t\rangle.$$
Thus the states $\mid J_1,J_2,\alpha_1,\alpha_2\rangle$ are temporally stable. 
\begin{remark}
$\bullet$~Instead of defining the states as in (\ref{s64}), if we define them as (notice that the change will not affect the calculations preceding this remark; thereby the following class of vectors also forms a set of CS)
\begin{eqnarray}
&&\mid J_1,J_2,\alpha_1,\alpha_2\rangle=\N_1(J_1,J_2)^{-1}\sum_{m=0}^{\infty}\frac{J_{1}^{m/2}}{\sqrt{\rho_2(m)}}e^{-ie_m\alpha_1}\N_2(J_2,m)^{-1}\nonumber\\
&&\hspace{2cm}\times\sum_{n=0}^{\infty}\frac{J_{2}^{n/2}}{\sqrt{\rho_1(m,n)}}e^{-iE(m,n)\alpha_2}\phi_{m}\otimes\chi_n\label{s69}
\end{eqnarray}
we can have
$$e^{iH_0t}\mid J_1,J_2,\alpha_1,\alpha_2\rangle=\mid J_1,J_2,\alpha_1,\alpha_2+t\rangle.$$
Still we have the temporal stability, but only the second part of the states evolve with time. In this case, if we shift the spectrum so that $E(m,0)=0$ we can have an action identity in the following sense:
\begin{eqnarray*}
\langle J_1,J_2,\alpha_1,\alpha_2\mid H_0\mid J_1,J_2,\alpha_1,\alpha_2\rangle&=&\N_1(J_1,J_2)^{-2}\sum_{m=0}^{\infty}\frac{J_{1}^{m}}{\rho_2(m)}\N_2(J_2,m)^{-2}\\&&\hspace{3cm}\times\sum_{n=1}^{\infty}\frac{J_{2}^{m}E(m,n)}{\rho_1(m,n)}=J_2.
\end{eqnarray*}
$\bullet$~If we shift $e_m$ and $E(m,n)$ backward by $e_0$ and $E(m,0)$ we get $\widetilde{e}_m=e_m-e_0=2Bm$ and $\widetilde{E}(m,n)=E(m,n)-E(m,0)=\pi^2 n(n+2)/d^2$, which is the same case considered in Remark \ref{re6}.
\end{remark}
\section{Dynamical algebra}
In this section we discuss the dynamical algebra associated to each set of temporally
 stable states of the previous sections. Here we follow the operator structure
 developed in Section \ref{GKCS}. That is, we follow the annihilation, creation 
and the number operators of (\ref{s45}).
\subsection{For the states of section \ref{NFIX}}\label{sub1}
When $n$ is fixed the spectrum $E(m,n)=B(2m+1)+\left(\frac{\pi (n+1)}{d}\right)^2$ 
can be written as
$$\overline{E}(m)=b_1 m+c_1,$$
where $b_1=2B$ and $c_1=B+\left(\frac{\pi (n+1)}{d}\right)^2$ are constants.
 The corresponding generators take the form (\ref{s45}) with $x_{m}=\overline{E}(m)$.
From (\ref{s46}) the commutators take the form
$$[\an,\an^{\dagger}]=b_1I,\;\;\;[\n,\an]=b_1\an^{\dagger},\;\;\;[\n,\an]=-b_1\an.$$
 Thus the dynamical algebra is isomorphic to the Weyl-Heisenberg algebra, 
$\mathfrak{g}_{\text{w-h}}$. To get the exact commutation relations of the 
Weyl-Heisenberg algebra one can define a new set of operators as follows.
$$\overline{\an}=\frac{1}{\sqrt{b_1}}\an,\;\;\;\overline{\an}^{\dagger}
=\frac{1}{\sqrt{b_1}}\an^{\dagger},\;\;\;\overline{\n}=\frac{1}{\sqrt{b_1}}\n.$$
In terms of these new operators one gets,
$$[\overline{\an},\overline{\an}^{\dagger}]=I,\;\;\;
[\overline{\n},\overline{\an}^{\dagger}]=\overline{\an}^{\dagger},\;\;\;
[\overline{\n},\overline{\an}]=-\overline{\an}.$$
\subsection{For the states of section \ref{MFIX}}\label{sub2}
When $m$ is fixed the spectrum $E(m,n)=B(2m+1)+\displaystyle\left(\frac{\pi (n+1)}{d}\right)^2$
 can be written as
$$\widetilde{E}(n)=b_2 (n+1)^2+c_2,$$
where $b_2=\pi^2/d^2$ and $c_2=B(m+1)$ are constants. The corresponding generators take the form 
(\ref{s45}) with $e_{n}=\widetilde{E}(n)$. Let us see the commutation relations. 
It can be easily seen that
$$[\an,\an^{\dagger}]\psi_{m,n}=b_2(2n+3)\psi_{m,n}.$$
As it was done in \cite{Tem}, let us define a new set of operators
\begin{equation}\label{s71}
\overline{\an}=\frac{1}{\sqrt{b_2}}\an,\;\;\;
\overline{\an}^{\dagger}=\frac{1}{\sqrt{b_2}}\an^{\dagger},\;\;\;
\overline{\n}\psi_{m,n}=(n+\frac{3}{2})\psi_{m,n}.
\end{equation}
With these new operators we obtain
\begin{equation}\label{s72}
[\overline{\an},\overline{\an}^{\dagger}]=2\overline{\n},\;\;\;[\overline{\n},\overline{\an}]=-\overline{\an},\;\;\;[\overline{\n},\overline{\an}^{\dagger}]=\overline{\an}^{\dagger}.
\end{equation}
The above commutation relations are  the ones satisfied by the generators 
of the algebra $\mathfrak{su(1,1)}$ of the classical group $SU(1,1)$. 
Thus in this case the dynamical algebra is isomorphic to $\mathfrak{su(1,1)}$.
\subsection{For the states of section \ref{PTEN}}\label{sub3}
Since $\rho_1(m)=e_m!$, $\rho_2(n)=\epsilon_n!$, $e_m=B(2m+1)$ and $\epsilon_n=\displaystyle{\frac{\pi^2(n+1)^2}{d^2}}$, let us define two sets of operators as follows:
\begin{eqnarray}
\an_1\phi_m&=&\sqrt{e_m}\phi_{m-1},\;\;\an_{1}^{\dagger}\phi_{m}=\sqrt{e_{m+1}}\phi_{m+1},\;\;\n_1\phi_{m}=e_m\phi_m\label{s73}\\
\an_2\chi_n&=&\sqrt{\epsilon_n}\chi_{n-1},\;\;\an_{2}^{\dagger}\chi_{n}=\sqrt{\epsilon_{n+1}}\chi_{n+1},\;\;\n_2\chi_{n}=\epsilon_n\chi_n.\label{s74}
\end{eqnarray}
For the operators $\an_1,\an_1^{\dagger},\n_1$, the commutators take the form,
$$[\an_1,\an_1^{\dagger}]=2BI,\;\;\;[\n_1,\an_1^{\dagger}]=2B\an_1^{\dagger},\;\;\;[\n_1,\an_1]=-2B\an_1.$$
Thus the dynamical algebra is isomorphic to  $\mathfrak{g}_{\text{w-h}}$. To get the exact commutation relations of  $\mathfrak{g}_{\text{w-h}}$ one can define a new set of operators as follows.
$$\overline{\an}_1=\frac{1}{\sqrt{2B}}\an_1,\;\;\;\overline{\an}_{1}^{\dagger}=\frac{1}{\sqrt{2B}}\an_1^{\dagger},\;\;\;\overline{\n}_1=\frac{1}{\sqrt{2B}}\n_1.$$
In terms of these new operators one gets,
$$[\overline{\an}_1,\overline{\an}_{1}^{\dagger}]=I,\;\;\;[\overline{\n}_1,\overline{\an}_{1}^{\dagger}]=\overline{\an}_{1}^{\dagger},\;\;\;[\overline{\n}_1,\overline{\an}_1]=-\overline{\an}_1.$$
For the operators $\an_2,\an_{2}^{\dagger}$ we get
$$[\an_2,\an_2^{\dagger}]\chi_{n}=\frac{2\pi^2}{d^2}(n+\frac{3}{2})\chi_{n}.$$
By defining a new set of operators
\begin{equation}\label{s75}
\overline{\an}_2=\frac{d}{\pi}\an_2,\;\;\;\overline{\an}_2^{\dagger}=\frac{d}{\pi}\an_2^{\dagger},\;\;\;\overline{\n}_2\chi_{n}=(n+\frac{3}{2})\chi_{n}
\end{equation}
it can readily be seen that the commutators take the following form.
\begin{equation}\label{s76}
[\overline{\an}_2,\overline{\an}_2^{\dagger}]=2\overline{\n}_2,\;\;\;[\overline{\n}_2,\overline{\an}_2]=-\overline{\an}_2,\;\;\;[\overline{\n}_2,\overline{\an}_2^{\dagger}]=\overline{\an}_2^{\dagger},
\end{equation}
which are  the commutation relations satisfied by the generators of the algebra $\mathfrak{su(1,1)}.$
Now for the set of CS we define the following set of operators,
\begin{equation}\label{s77}
\an=\an_1\otimes \an_2,\;\;\;\an^{\dagger}=\an_{1}^{\dagger}\otimes \an_{2}^{\dagger},\;\;\;\n=\n_1\otimes \overline{\n}_{2},
\end{equation}
 Thus the algebra associated to the CS is isomorphic to the tensor product of the two algebras, $\mathfrak{g}_{\text{w-h}}$ and $\mathfrak{su(1,1)}$ , that is, $\mathfrak{g}_{\text{w-h}}\otimes \mathfrak{su(1,1)}$.
If we take the operators as
\begin{equation}\label{s78}
\overline{\an}=\overline{\an}_1\otimes \overline{\an}_2,\;\;\;\overline{\an}^{\dagger}=\overline{\an}_{1}^{\dagger}\otimes \overline{\an}_{2}^{\dagger},\;\;\;\overline{\n}=\overline{\n}_1\otimes \overline{\n}_{2},
\end{equation}
we get the exact commutation relations of $\mathfrak{g}_{\text{w-h}}\otimes \mathfrak{su(1,1)}$. One can also define another set of operators as follows.
\begin{eqnarray}\label{s79}
\mathbf{a}\phi_m\otimes\chi_n&=&\sqrt{e_m\epsilon_n}\phi_{m-1}\otimes\chi_{n-1}\nonumber,\;\;\;\mathbf{a}\phi_0\otimes\chi_0=0\\
\mathbf{a}^{\dagger}\phi_m\otimes\chi_n&=&\sqrt{e_{m+1}\epsilon_{n+1}}\phi_{m+1}\otimes\chi_{n+1}\\
\mathbf{n}\phi_m\otimes\chi_n&=&e_m\epsilon_n\phi_{m}\otimes\chi_{n}\nonumber
\end{eqnarray}
Observe that here also the CS become the eigenstates of $\mathbf{a}$. But it may be difficult to identify this algebra to a known type.
\subsection{For the states of section \ref{SUMDEP}}\label{sub4}
Since $\rho_1(m)=e_m!$, $\rho_2(n)=E(m,n)!$, let us define two sets of operators as follows:
\begin{eqnarray*}
\an_1\phi_m&=&\sqrt{e_m}\phi_{m-1},\;\;\an_{1}^{\dagger}\phi_{m}=\sqrt{e_{m+1}}\phi_{m+1},\;\;\n_1\phi_{m}=e_m\phi_m\label{set1}\\
\an_2\chi_n&=&\sqrt{E(m,n)}\chi_{n-1},\;\;\an_{2}^{\dagger}\chi_{n}=\sqrt{E(m,n+1)}\chi_{n+1},\;\;\an_2\chi_{n}=E(m,n)\chi_n\label{set2}
\end{eqnarray*}
Again an analogue of subsection \ref{sub1} can be worked out for the operators $\an_1,\an_{1}^{\dagger},\n_1$. Thus the operators generate the algebra $\mathfrak{g}_{\text{w-h}}$. Since within the second sum of the CS $m$ is considered as a constant we are in the exact situation of subsection \ref{sub2}. Thus the algebra generated by $\an_2,\an_{2}^{\dagger},\n_2$  is isomorphic to the algebra $\mathfrak{su(1,1)}$. The rest of the details follows from the subsection \ref{sub3}.
\section{Statistical quantities}
Quantum revivals are associated to wave functions. A revival of a
 wave function occurs when a wave function evolve in time to a state closely
 reproducing its initial form. Further the weighting distribution is crucial 
for understanding the temporal behavior of the wave function. In the case of the
 states (\ref{s41}), the probability of finding the state $\eta_m$ in the state 
$\mid J,\alpha\rangle$ is given by
$$P(m,J)=\abs{\langle \eta_m\mid J,\alpha\rangle}^2.$$
A quantitative estimate is given by the so-called Mandel parameter,
$$Q=\frac{\langle J,\alpha\mid \n^2\mid J,\alpha\rangle-\langle J,\alpha\mid \n\mid J,
\alpha\rangle^2-\langle J,\alpha\mid \n\mid J,\alpha\rangle}{\langle J,\alpha\mid \n\mid J,
\alpha\rangle}$$
where $\n\eta_m=e_m\eta_m$. If the Photon distribution is Poissonian then $Q=0$. 
If $Q<0$ it is called sub-Poissonian and if $Q>0$ it is called super-Poissonian 
\cite{Tem}. In this section we explicitely calculate the weighting distribution and 
the Mandel parameter for each of the CS discussed in the above sections.
\subsection{For the states of Eq.(\ref{s52})}
For this class of states we obtain
$$P(m,J)=\frac{\abs{J}^m}{\mathcal N(J,n)^2\rho(m)}=\frac{\left(J/2B\right)^m}
{{}_1F_{1}(1;\gamma;\frac{J}{2B})(\gamma)_m}.$$
Since $\n\mid\psi_{m,n}\rangle=E(m,n)\mid\psi_{m,n}\rangle$ and $E(0,n)\not=0$ we have
\begin{eqnarray*}
\n\mid J,\alpha,n\rangle&=&\N(J,n)^{-1}\sum_{m=0}^{\infty}\frac{J^{m/2}E(m,n)}
{\sqrt{\rho(m)}}e^{-iE(m,n)\alpha}\mid \psi_{m,n}\rangle
\end{eqnarray*}
and
\begin{eqnarray*}
\n^2\mid J,\alpha,n\rangle&=&\N(J,n)^{-1}\sum_{m=0}^{\infty}\frac{J^{m/2}E(m,n)^2}
{\sqrt{\rho(m)}}e^{-iE(m,n)\alpha}\mid \psi_{m,n}\rangle.
\end{eqnarray*}
Thus
\begin{eqnarray*}
\langle J,\alpha,n\mid \n\mid J,\alpha,n\rangle&=&\frac{J}{\gamma}
\frac{{}_1F_1(2;1+\gamma;\frac{J}{2B})}{{}_1F_1(1;\gamma;\frac{J}{2B})}+\omega
\end{eqnarray*}
where $\omega=B+\left(\frac{\pi(n+1)}{d}\right)^2$ and thereby $E(m,n)=2Bm+\omega$.
Further
\begin{eqnarray*}
&&\langle J,\alpha,n\mid \n^2\mid J,\alpha,n\rangle\\
&&\hspace{0.5cm}=\frac{1}{{}_1F_{1}(1;\gamma;\frac{J}{2B})}\big[\frac{2J(B+\omega)}
{\gamma}{}_1F_1(2;\gamma+1;\frac{J}{2B})+\frac{2J^2}{\gamma(\gamma+1)}
{}_1F_1(3;\gamma+2;\frac{J}{2B})\big]+\omega^2.
\end{eqnarray*}
Therefore
\begin{eqnarray*}
Q&=& \frac{2J(B+\omega)(\gamma+1){}_1F_1(2;\gamma+1;\frac{J}{2B})+2J^2
{}_1F_1(3;\gamma+2;\frac{J}{2B})+\gamma(\gamma+1)\omega^2{}_1F_1(1;\gamma;\frac{J}{2B})}
{(\gamma+1)[J{}_1F_1(2;\gamma+1;\frac{J}{2B})+\gamma\omega{}_1F_1(1;\gamma;\frac{J}{2B})]}\\
&&-\frac{J{}_1F_1(2;\gamma+1;\frac{J}{2B})}{\gamma{}_1F_1(1;\gamma;\frac{J}{2B})}-\omega-1.
\end{eqnarray*}
For particular values of $B,d$ and $n$ the sign of $Q$ can be determined.
\subsection{For the states of Eq.(\ref{s56})}
We have
$$P(n,J)=\frac{d^{2n}J^n}{\pi^{2n}(\beta)_n(\overline\beta)_n{}_1F_2(1;\beta,\overline{\beta};
\frac{d^2J}{\pi^2})}.$$
For fixed $m$, $E(m,n)=p+q(n+1)^2$ where $p=B(2m+1)$ and $q=\pi^2/d^2$. Since 
$E(m,0)\not=0$ and $\n\mid\psi_{mn}\rangle=E(m,n)\mid\psi_{mn}\rangle$ we have
\begin{eqnarray*}
&&\langle J,\alpha,m\mid \n\mid J,\alpha,m\rangle=p+\frac{q}{\N(J,m)^2}
\sum_{n=0}^{\infty}\frac{J^n(n+1)^2}{\rho(n)}=p+qQ_{1}
\end{eqnarray*}
where
$$Q_1=\frac{1}{{}_1F_{2}(1;\overline{\beta},\beta;\frac{d^2J}{\pi^2})}
\left[{}_1F_{2}(2;\overline{\beta},\beta;\frac{d^2J}{\pi^2})+\frac{2Jd^2}
{\abs{\beta}^2\pi^2}{}_1F_{2}(3;\overline{\beta}+1,\beta+1;\frac{d^2J}{\pi^2})\right]$$
and
\begin{eqnarray*}
&&\langle J,\alpha,m\mid \n^2\mid J,\alpha,m\rangle
=\N(J,m)^{-2}\sum_{n=0}^{\infty}\frac{J^{n}d^{2n}(p+q(n+1)^2)^2}{\pi^{2n}
(\beta)_n(\overline{\beta})_n}=p^2+2pqQ_1+q^2Q_2
\end{eqnarray*}
where 
\begin{eqnarray*}
Q_2&=&\frac{1}{{}_1F_{2}(1;\overline{\beta},\beta;\frac{d^2J}{\pi^2})}\big[\frac{2Jd^2+\pi^2}
{\pi^2}{}_1F_{2}(2;\overline{\beta},\beta;\frac{d^2J}{\pi^2})\\&&-\frac{2Jd^2(\pi^2\abs{\beta}^2
-Jd^2-7\pi^2)}{\abs{\beta}^2\pi^4}{}_1F_{2}(3;\overline{\beta}+1,\beta+1;\frac{d^2J}{\pi^2})\\
&&-\frac{6J^2d^4(\beta+\overline{\beta}-5)}{\abs{\beta}^2\pi^4(\beta+1)(\overline{\beta}+1)}
{}_1F_{2}(4;\overline{\beta}+2,\beta+2;\frac{d^2J}{\pi^2})\big]
\end{eqnarray*}
Thus
$$Q=\frac{p^2+2pqQ_1+q^2Q_2}{p+qQ_1}-p-qQ_1-1.$$
Here again for specific values of $B,d$ and $m$ the sign of $Q$ can be determined.
\subsection{For the states of Eq.(\ref{s61})}
The probability of finding the state $\phi_m\otimes\chi_n$ in the state 
$\mid J_1,J_2,\alpha_1,\alpha_2\rangle$ is given by
$$P(m,n,J_1,J_2)=\frac{J_{1}^{m}J_{2}^{n}}{{}_1F_{1}(1;\frac{3}{2};
\frac{J_1}{2B}){}_0F_1(-;1;\frac{J_2d^2}{\pi^2})2^mB^m\left(\frac{\pi}{d}\right)^{2n}
(2)_n(2)_n\left(\frac{3}{2}\right)_m}$$
Since $\n_1\phi_m=e_m\phi_m$,\;$\n_2\chi_n=\epsilon_n\chi_n$,
 $\n=\n_1\otimes \n_2$, and $e_0\not=0,\epsilon_0\not=0$ we have
$$\langle J_1,J_2,\alpha_1,\alpha_2\mid \n\mid J_1,J_2,\alpha_1,\alpha_2\rangle
=\frac{B\pi^2Q_4}{d^2}[2Q_3+1]$$
where
$$Q_3=\frac{1}{\N_1(J_1)^2}\sum_{m=0}^{\infty}\frac{J_1^mm}{2^mB^m(\frac{3}{2})_m}
=\frac{J}{3B\N_1(J_1)^2}{}_1F_1(2;\frac{5}{2};\frac{J}{2B})$$
and
$$Q_4=\frac{1}{\N_2(J_2)^2}\sum_{m=0}^{\infty}\frac{d^{2m}J_2^m(m+1)^2}{\pi^{2m}(2)_m(2)_m}.$$
Further
\begin{eqnarray*}
\langle J_1,J_2,\alpha_1,\alpha_2\mid \n^2\mid J_1,J_2,\alpha_1,\alpha_2\rangle
&=&\langle J_1,J_2,\alpha_1,\alpha_2\mid \n_1^2\otimes \n_2^2\mid J_1,J_2,\alpha_1,\alpha_2
\rangle\\
&=&\N_{1}(J_1)^{-2}\sum_{m=0}^{\infty}\frac{J_{1}^{m}e_m^2}{e_m!}\N_{2}(J_2)^{-2}
\sum_{n=0}^{\infty}\frac{J_2^n\epsilon_n^2}{\epsilon_n!}\\
&=&\frac{B^2\pi^4Q_6}{d^4}[4Q_5+4Q_3+1]
\end{eqnarray*}
where
$$Q_5=\frac{1}{\N_1(J_1)^2}\sum_{m=0}^{\infty}\frac{J_1^mm^2}{2^mB^m(\frac{3}{2})_m}$$
and $Q_6$ can be obtained from $Q_2$ by substituting $J=J_2,\beta=\overline{\beta}=2$ 
and $\N(J,m)=\N_1(J_1)$ in the expression of $Q_2$. $Q_4$ can be obtained from $Q_1$ by
 the same substitution. Thereby we have
$$Q=\frac{B\pi^2Q_6(4Q_5+4Q_3+1)}{d^2Q_4(2Q_3+1)}-\frac{B\pi^2Q_4(2Q_3+1)}{d^2}-1.$$
For specific values of $B$ and $d$ the sign of $Q$ can be determined.
\subsection{For the states of Eq.(\ref{s64})}
The probability of finding the state $\phi_m\otimes\chi_n$ in the state $\mid J_1,J_2,
\alpha_1,\alpha_2\rangle$ is given by
$$P(m,n,J_1,J_2)=\frac{J_{1}^{m}J_{2}^{n}}{{}_1F_2(1;\beta,\overline{\beta};\frac{d^2J_2}
{\pi^2})\N_1(J_1,J_2)^22^mB^m\left(\frac{3}{2}\right)_n\left(\frac{\pi}{d}\right)^{2n}
(\beta)_n(\overline{\beta})_n}$$
Further, by taking $\n_1\phi_m=e_m\phi_m$, $\n_2\chi_n=E(m,n)\chi_n$
 and $\n=\n_1\otimes \n_2$ one can find $Q$ as in the previous section. 
Since we do not have a closed form for $\N_1(J_1,J_2)$ we avoid calculating it.

\section{conclusion}
Eigenfunctions and eigenvalues of the free magnetic Sch\"odinger operator 
$H_0=\frac{1}{2M}(\text{\bf P}-\frac{e}{c}\text{\bf A})^2$ were discussed. 
The eigenfunctions were realized as an orthonormal basis of a Hilbert space.
 Four classes of temporally stable CS associated to the eigenfunctions and eigenvalues
 of the operator $H_0$ were demonstrated. The first two classes were constructed with one degree
 of freedom and the last two with two degrees of freedom. To each class of CS 
the corresponding dynamical algebra was specified. The dynamical algebras were identified to 
the Weyl-Heisenberg algebra, $\mathfrak{su(1,1)}$ algebra and their tensor products. 
For each class of CS, quantum statistical quantities were calculated explicitely.

\section*{acknowledgment}
The authors would like to thank G. Honnouvo for suggesting the problem of this paper. 
Partial financial support of this work under Grant No. GP249507 from the Natural Sciences
 and Engineering Research Council of Canada is gratefully acknowledged [NS].


\begin{thebibliography}{XXXX}
\bibitem{Ali} Ali~S.T., Antoine~J-P. and Gazeau~J-P., {\em Coherent States, Wavelets and Their Generalizations}, Springer, New York (2000).
\bibitem{Tem} Antoine~J-P., Gazeau~J-P., Monceau~P., Klauder~J. R. and Penson~K. A., ``Temporally stable coherent states for infinite well and P\"oschl-Teller potentials", J.~Math.~Phys. {\bf 42}, 2349-2387 (2001).
\bibitem{D} Doetsch~G., {\em Handbuch der Laplace-Transformation}, Band 1, Birkh\"auser Verlag Basel (1971), Theorems 4 and 5, pp. 74-79.
\bibitem{key10} Erd\'elyi A.,  Magnus W.,  Oberhettinger F. and  Tricomi F.G., {\it Higer Transcendental functions,} McGraw-Hill,  New York (1953).
\bibitem{key3} Erd\'elyi~A.,  Magnus~W.,  Oberhettinger~F. and  Tricomi~F.G., {\it Tables of integral transforms}, McGraw-Hill,  New York (1953).
\bibitem{Ex} Exner~P. and Nemcova~K., ``Magnetic layers with periodic point perturbations", Rep.~Math.~Phys. {\bf 52}, 255-280 (2003).
\bibitem{Exn} Exner~P. and Nemcova~K., ``Quantum mechanics of layers with a finite number of point perturbations", J.~Math.~Phys. {\bf 43}, 1152-1184 (2002).
\bibitem{F} Fakhri~H., ``Generalized Klauder-Perelomov and Gazeau-Klauder coherent states for Landau levels", Phys.~Lett.~A, {\bf 313}, 243-251 (2003).
\bibitem{Gk} Gazeau~J-P. and Klauder~J.R., ``Coherent states for systems with discrete and continuous spectrum", J.~Phys.~A: Math.~Gen. {\bf 32}, 123-132 (1999).
\bibitem{GJ} Gazeau~J-P., Hsian~Y. and Jellal~A., ``Exact trace formulas for two-dimensional electron magnetism", Phys.~Rev.~B {\bf 65}, 094427 (2002). 
\bibitem{K-hyd} Klauder~J.R., ``Coherent states for the hydrogen atom", J.~Phys.~A {\bf 29}, L293-L298 (1996).
\bibitem{Gla} Klauder~J.R and Skagerstam~B.S, {\em Coherent States, Applications in Physics and Mathematical Physics}, World Scientific, Singapore (1985).
\bibitem{K} Kondratiev~V. and Shubin~M., ``Discreteness of spectrum for the magnetic Schr\"odinger operators", Comm.~Partial Differential Equations {\bf 27}, 477-525 (2002).
\bibitem{M} Marichev~O.I., {\em Higher Transcendental Functions, Theory and Algorithmic Tables}, Ellis Harwood, Chichester (1983).
\bibitem{Nay} Naylor~D., ``An asymptotic expansion of the Kontorovich-Lebedev transform of damped oscillatory functions", J.~Comp.~App.~Math. {\bf 145}, 21-30 (2002).
\bibitem{Ng} Novaes~M. and Gazeau~J-P., ``Multidimensional generalized coherent states", J.~Phys.~A: Math.~Gen., {\bf 36}, 199-212 (2003).
\bibitem{Pre} P\'er\'elomov~A.M.,
{\em Generalized Coherent States and Their Applications}, Springer-Verlag, Berlin (1986).
\bibitem{H} Richard~L. Hall, Nasser Saad and Attila B. von Keviczky, ``Matrix elements for a generalized spiked harmonic oscillator", J.~Math.~Phys. {\bf 39}, 6345-6352 (1998).
\bibitem{HNA} Richard~L. Hall, Nasser Saad and Attila B. von Keviczky, ``Spiked harmonic oscillators", J.~Math.~Phys. {\bf 43}, 94-112 (2002).
\bibitem{R} Rudin~W., {\em Real and Complex Analysis}, 3$^\text{rd}$, McGraw-Hill, New York (1987), Theorem 1.34, p.26.
\bibitem{S} Smithies~F., {\em Integral Equations}, Cambridge University Press, Cambridge (1970), p. 12.
\bibitem{W} Weidmann~J., {\em Linear Operators in Hilbert Spaces}, Springer-Verlag, New York (1980).




\end{thebibliography}
\end{document}